\def\ie{{\it i.e.}}

\def\~{{$\tilde{\phantom{a}}$}}

\documentclass [12pt] {article}
\usepackage{epsfig}
\usepackage{color}

\def\thebibliography#1{\section{References}\markboth
 {REFERENCES}{REFERENCES}\list
 {[\arabic{enumi}]}{\settowidth\labelwidth{[#1]}\leftmargin\labelwidth
 \advance\leftmargin\labelsep
 \usecounter{enumi}}
 \def\newblock{\hskip .11em plus .33em minus -.07em}
 \sloppy
 \sfcode`\.=1000\relax}
\def\upcite#1{\raise6pt\hbox{\scriptsize
\cite{#1}}}
\def\lsim{\mathrel {\vcenter {\baselineskip 0pt \kern 0pt
    \hbox{$<$} \kern 0pt \hbox{$\sim$} }}}
\def\gsim{\mathrel {\vcenter {\baselineskip 0pt \kern 0pt
    \hbox{$>$} \kern 0pt \hbox{$\sim$} }}}
\def\gtlt{\mathrel {\vcenter {\baselineskip 0pt \kern 0pt
    \hbox{$>$} \kern 0pt \hbox{$<$} }}}

\def\hline{\noalign{\hrule \vskip2pt}}

\def\|{\ifmmode\Vert\else \char`\|\fi}
\ifx\oldzeta\undefined                          
  \let\oldzeta=\zeta                            
  \def\zzeta{{\raise 2pt\hbox{$\oldzeta$}}}     
  \let\zeta=\zzeta                              
\fi

\ifx\oldchi\undefined                           
  \let\oldchi=\chi                              
  \def\cchi{{\raise 2pt\hbox{$\oldchi$}}}       
  \let\chi=\cchi                                
\fi

\def\frac#1#2{{#1 \over #2}}

\def\half{\ifinner {\scriptstyle {1 \over 2}}
   \else {1 \over 2} \fi}

\def\simge{\mathrel{%
   \rlap{\raise 0.511ex \hbox{$>$}}{\lower 0.511ex \hbox{$\sim$}}}}
\def\simle{\mathrel{
   \rlap{\raise 0.511ex \hbox{$<$}}{\lower 0.511ex \hbox{$\sim$}}}}

\def\buildchar#1#2#3{{\null\!                   
   \mathop#1\limits^{#2}_{#3}                   
   \!\null}}                                    
\def\overcirc#1{\buildchar{#1}{\circ}{}}

\def\slashchar#1{\setbox0=\hbox{$#1$}           
   \dimen0=\wd0                                 
   \setbox1=\hbox{/} \dimen1=\wd1               
   \ifdim\dimen0>\dimen1                        
      \rlap{\hbox to \dimen0{\hfil/\hfil}}      
      #1                                        
   \else                                        
      \rlap{\hbox to \dimen1{\hfil$#1$\hfil}}   
      /                                         
   \fi}                                         %

\def\subrightarrow#1{
  \setbox0=\hbox{
    $\displaystyle\mathop{}
    \limits_{#1}$}
  \dimen0=\wd0
  \advance \dimen0 by .5em
  \mathrel{
    \mathop{\hbox to \dimen0{\rightarrowfill}}
       \limits_{#1}}}                           

\def\overlay#1#2{\ifmmode%
\setbox0=\hbox{$#1$}%
\setbox1=\hbox to\wd0{\hss$#2$\hss}\else%
\setbox0=\hbox{#1}%
\setbox1=\hbox to\wd0{\hss#2\hss}\fi%
#1\hskip-\wd0\box1 }

\def\pmb#1{\leavevmode\setbox0=\hbox{#1}%
\kern-.02em\copy0\kern-\wd0
\kern.04em\copy0\kern-\wd0
\kern-.02em\raise.04em\box0 }

\def\vereq#1#2{\lower3pt\vbox{\baselineskip1.5pt \lineskip1.5pt
\ialign{$\m@th#1\hfill##\hfil$\crcr#2\crcr\sim\crcr}}}

\def\tensor#1{\protect\@ontopof{#1}{\leftrightarrow}{1.15}\mathord{\box2}}
\def\overstar#1{\protect\@ontopof{#1}{\ast}{1.15}\mathord{\box2}}
\def\overdots#1{\protect\@ontopof{#1}{\cdots}{1.0}\mathord{\box2}}
\def\overcirc#1{\protect\@ontopof{#1}{\circ}{1.2}\mathord{\box2}}
\def\loarrow#1{\protect\@ontopof{#1}{\leftarrow}{1.15}\mathord{\box2}}
\def\roarrow#1{\protect\@ontopof{#1}{\rightarrow}{1.15}\mathord{\box2}}

\def\@ontopof#1#2#3{%
{\mathchoice
{\@@ontopof{#1}{#2}{#3}\displaystyle\scriptstyle}%
{\@@ontopof{#1}{#2}{#3}\textstyle\scriptstyle}%
{\@@ontopof{#1}{#2}{#3}\scriptstyle\scriptscriptstyle}%
{\@@ontopof{#1}{#2}{#3}\scriptscriptstyle\scriptscriptstyle}%
}%
}

\def\@@ontopof#1#2#3#4#5{%
\setbox0=\hbox{$#4#1$}%
\setbox1=\hbox{$#5#2$}%
\setbox2=\hbox{}\ht2=\ht0 \dp2=\dp0 %
\ifdim\wd0>\wd1 %
\setbox1=\hbox to\wd0{\hss\box1\hss}%
\mathord{\rlap{\raise#3\ht0\box1}\box0}%
\else   %
\setbox1=\hbox to.9\wd1{\hss\box1\hss}%
\setbox0=\hbox to\wd1{\hss$#4\relax#1$\hss}%
\mathord{\rlap{\copy0}\raise#3\ht0\box1}%
\fi
}%

\def\lambdabar{\protect\@lambdabar}
\def\@lambdabar{%
\relax
\bgroup
\def\@tempa{\hbox{\raise.73\ht0
\hbox to0pt{\kern.25\wd0\vrule width.5\wd0
height.1pt depth.1pt\hss}\box0}}%
\mathchoice{\setbox0\hbox{$\displaystyle\lambda$}\@tempa}%
{\setbox0\hbox{$\textstyle\lambda$}\@tempa}%
{\setbox0\hbox{$\scriptstyle\lambda$}\@tempa}%
{\setbox0\hbox{$\scriptscriptstyle\lambda$}\@tempa}%
\egroup
}

\def\corresponds{{\lower.2ex\hbox{=}}{\rm\kern-.75em^\triangle}}
\def\succsim{\succ\kern-.9em_\sim\kern.3em}
\def\precsim{\prec\kern-1em_\sim\kern.3em}
\def\slantfrac#1#2{\kern1em^{#1}\kern-.3em/\kern-.1em_{#2}}

\begin{document}
                                                                
\begin{center}
{\Large\bf Hidden Momentum in a Coaxial Cable}
\\

\medskip

Kirk T.~McDonald
\\
{\sl Joseph Henry Laboratories, Princeton University, Princeton, NJ 08544}
\\
(March 28, 2002)
\end{center}

\section{Problem}

Calculate the electromagnetic momentum and identify the ``hidden'' mechanical
momentum in a coaxial cable of length $L$, inner radius $a$, outer radius $b$, when
a battery is connected to one end and a load resistor $R_0$ is connected to the other.
The current may be taken as uniformly distributed over the inner conductor, 
which has resistivity $\rho$.  The outer conductor has negligible resistivity, and
the current flows on it in a thin sheet at radius $b$.

\section{Solution}

This problem is based on sec.~17 of \cite{Sommerfeld}, and on prob.~7.57, ex.~8.3
and ex.~12.12 of \cite{Griffiths}.

We denote the resistance per unit length along inner conductor as
\begin{equation}
R = {\rho \over \pi a^2}\, .
\label{s1}
\end{equation}
Then, the total resistance of the cable plus load resistor is $R_0 + R L$.  To have
current $I$ in the system, the battery must have voltage
\begin{equation}
V = I (R_0 + R L).
\label{s2}
\end{equation}
The current $I$ causes a magnetic field that is readily calculated via Amp\`ere's law
to be (in Gaussian units, and in a cylindrical coordinate system $(r,\phi,z)$ with
the coaxial cable centered on the $z$ axis),
\begin{equation}
{\bf B} = {2 I \over c} \hat\phi \left\{ \begin{array}{ll}
{r \over a^2}  & (r < a), \\
{1 \over r} & (a < r < b), \\
0 & (r > b).
\end{array}
\right.
\label{s3}
\end{equation}
Inside the wire the electric field is ${\bf E}(r<a) = I R \hat{\bf z}$, as needed
to drive the current $I$ against the resistivity $\rho$.  Since the tangential
component of the electric field is continuous across a boundary, there must be
some electric field in the region $r > a$ as well.  Indeed, a charge distribution
$Q(z)$ is needed on the surface of the inner conductor to shape the interior
electric field to be purely longitudinal.

An analysis of the electric field
can be based on the convention that the electric potential $V(r,z)$
is equal to zero on the outer conductor, and is also zero
on the plane $z = 0$ (which is not necessarily
inside the wire of length $L$).  That is, we suppose the cable extends from
$z = - L - R_0 / R$ (the position of the battery) to $z = - R_0 / R$ (the position
of the resistor), so that the electric potential for $r \leq a$ 
can be written as
\begin{equation}
V(r \leq a,z) = - I R z.
\label{s4}
\end{equation}
Thus, the potential of the inner conductor at the position of the load resistor is
$I R_0$, and the potential at the position of the battery is $I R (L + R_0/R)$, 
\ie, the battery voltage (\ref{s2}).

The capacitance per unit length between the inner and outer conductors of the
coaxial cable is well known to be
\begin{equation}
C = {1 \over 2 \ln (b/a)}\, .
\label{s5}
\end{equation}
The charge $Q(z)$ per unit length on the inner conductor is therefore
\begin{equation}
Q(z) = C V(r=a,z) = - {I R z \over 2 \ln (b/a)} = {I R z \over 2 \ln (a/b)}\, ,
\label{s6}
\end{equation}
assuming that $L \gg b$ so that $Q(z)$ is essentially constant over length 
$\Delta z \ll b$.  Further, the potential in the region $a < r < b$ is essentially
that for a long wire of charge density $Q(z)$, matched to the condition that
$V(r=b) = 0$, namely
\begin{equation}
V(a<r<b,z) = - 2 Q(z) \ln(r/b) = - {I R z \ln(r/b) \over \ln (a/b)}\, ,
\label{s7}
\end{equation}
which also matches eq.~(\ref{s4}) at $r = a$.  The potential (\ref{s7}) can also
be obtained by a separation-of-variables solution to Laplace's equation
\cite{Sommerfeld}.

The electric field is obtained by taking the gradient of eq.~(\ref{s7}), and we find
\begin{equation}
{\bf E} = I R \left\{ \begin{array}{ll}
\hat{\bf z} & (r < a), \\
{\ln(r/b) \over \ln(a/b)} \hat{\bf z} + {z \over r \ln(a/b)} \hat{\bf r}
   & (a < r < b), \\
0 & (r > b).
\end{array}
\right.
\label{s8}
\end{equation}

The electromagnetic momentum density is
\begin{equation}
{\bf p}_{\rm EM} = {{\bf S} \over c^2} = {{\bf E} \times {\bf B} \over 4 \pi c}
 = {I^2 R \over 2 \pi c^2} \left\{ \begin{array}{ll}
- {r \over a^2} \hat{\bf r} & (r < a), \\
- {\ln(r/b) \over r \ln(a/b)} \hat{\bf r}  
+ {z \over r^2 \ln(a/b)} \hat{\bf z}    & (a < r < b), \\
0 & (r > b).
\end{array}
\right.
\label{s9}
\end{equation}
The Poynting vector {\bf S} quantifies the flow of energy from the battery in
the region $(a < r < b,z = - L - R_0/R)$ to the inner conductor and to the
load resistor, where the energy is dissipated in Joule heating.

The total electromagnetic momentum in the cable is
\begin{eqnarray}
{\bf P}_{\rm EM} & = & \int {\bf p}_{\rm EM}\ d{\rm Vol}
 = {I^2 R\ \hat{\bf z} \over 2 \pi c^2 \ln(a/b)} \int_a^b 2 \pi r\ dr
\int_{-L - R_0/R}^{-R_0/R} dz {z \over r^2 } 
\nonumber \\
& = & {I^2 R L (L + 2 R_0 / R)  \over 2  c^2} \hat{\bf z}.
\label{s10}
\end{eqnarray}

We expect the total momentum of the system to be zero, as its center of mass is
at rest, even though there is internal motion associated with the electrical current.
A small amount of mechanical momentum is ``hidden'' in the conduction electrons
because the ratio of mechanical momentum density to current density is 
$\gamma m_e / e$, where $\gamma = 1 / \sqrt{1 - v^2/c^2}$ is velocity dependent, 
and the velocity of the electrons
is higher in the inner conductor than in the outer \cite{Griffiths,Calkin}.
The factor of $c^2$ in the denominator of eq.~(\ref{s10}) alerts us to the
relativistic origin of the mechanical momentum that opposes the electromagnetic
momentum.

It remains somewhat difficult to quantify the velocity difference of the conduction
electrons in different parts of the circuit.  The usual model of conduction is
that the drift velocity is constant in regions of constant electric field, so we
are led to state that the (average) velocity of all conduction electrons in
the inner conduction is $- v_a \hat{\bf z}$, 
and that a single velocity $v_b \hat{\bf z}$ characterizes
the motion of electrons in the outer conductor.  The electric fields in the
battery and the load resistor vary with radius as $1/r$, and the battery 
provides an additional electromotive force, not describable by an ordinary electric
field, that is the source of energy to drive the electrons through the circuit.

Perhaps the most straightforward hypothesis is that as the electrons 
of charge $-e$ pass through
the battery, the electromotive force increases their energy by $e V > 0$. 
However, this does not result in a value for the mechanical momentum that is
equal and opposite the electromagnetic momentum, as is particularly clear for
the case of a cable that runs from $-L/2$ to $L/2$ with batteries of voltage
$IRL/2$ at both ends, for which ${\bf P}_{\rm EM} = 0$.

A prescription that achieves the desired goal, and is consistent with the
assumption that the velocities $v_a$ and $v_b$ are constant throughout their
respective conductors, is that the energy gain of the conduction electrons is
given by the potential difference between the centers of the inner and outer
conductors, \ie, by the average of the potential difference at the two ends.  This
prescription supposes that the energy transfer to the conduction electrons as
they move radially at one end of the cable is affected by the situation at the 
other end of the cable, no matter how distant.  While this is counterintuitive,
it is not contradictory for a steady-state situation.



Under the above assumption, the energy gain of the conduction electrons between
the inner and outer conductors implies that
\begin{equation}
m_e (\gamma_a - \gamma_b) c^2 = \Delta U 
= {e \over 2} [\Delta V(-L-R_0/R) + \Delta V(-R_0/R)]
= e I (R L / 2 + R_0).
\label{s11}
\end{equation}
Labeling the conduction electron density per unit length in the inner and outer
conductors as $n_a$ and $n_b$, we have
\begin{equation}
I = e n_a v_a = e n_b v_b,
\label{s12}
\end{equation}
and so the mechanical momentum in the current is
\begin{equation}
{\bf P}_{\rm mech}= (- m_e \gamma_a n_a L v_a  + m_e \gamma_b n_b L v_b) \hat{\bf z}
= - {m_e I L \over e} (\gamma_a - \gamma_b) \hat{\bf z}
= - {I^2 R L \over 2c^2} (L + 2 R_0 / R) \hat{\bf z}.
\label{s13}
\end{equation}
This cancels the electromagnetic momentum (\ref{s10}), and the total momentum
of the system (which is ``at rest'') is zero.


Another confirmation of the result (\ref{s10}) can be found by supposing the current
$I$ drops to zero with time.  The changing magnetic field induces a longitudinal
electric field that pushes on the charge on the surface of the inner conductor,
leading to a force on the wire.  The force on the conduction electrons merely slows
the decrease of the current, but does not cause a net force on the wire.  By
Faraday's law, the induced electric field at $r = a$ is
\begin{equation}
E_{z,\rm induced}(r=a) = - {1 \over c} {d \over dt} \int_a^b B_\phi\ dr
= - {2 \over c^2 a} {dI \over dt} \ln(b/a).
\label{s14}
\end{equation}
Note that $E_{z,\rm induced}(r=b) = 0$.
The additional force on the surface charge is
\begin{equation}
F_{z,\rm induced} = \int_{-L - R_0/R}^{-R_0/R} Q(z) E_{z,\rm induced}(r=a)\ dz 
= -{ R L (L + 2 R_0 / R)  \over 2  c^2} {dI^2 \over dt},
\label{s15}
\end{equation}
using eq.~(\ref{s6}).
The momentum kick to the wire as the current falls to zero is therefore
\begin{equation}
\Delta {\bf P}_{\rm mech} = \hat{\bf z} \int F_{z,\rm induced}\ dt
= {I^2 R L (L + 2 R_0 / R)  \over 2  c^2} \hat{\bf z} = {\bf P}_{\rm EM,initial}.
\label{s16}
\end{equation}
This result reinforces the interpretation of eq.~(\ref{s10}) as field momentum stored
in the system, that could be converted to mechanical momentum.  Of course,
as the current drops to zero, the ``hidden'' mechanical
momentum does also.  The total momentum is zero at all times, and the field
momentum is not transformed into net motion of the coaxial cable.  

Since the nonzero electromagnetic momentum of a coaxial cable at rest is
always canceled by the ``hidden'' mechanical momentum,  both of these entities
can be safely neglected by the pragmatic physicist in this case.  
Electromagnetic momentum
is of greater significance in dynamic phenomena, in which the mechanical momentum is
``evident'' rather than ``hidden'', and in which Newton's 3rd law for ``evident''
mechanical momentum is not satisfied
unless the electromagnetic momentum is taken into account \cite{trans}.

\end{document}